\documentclass[conference, 10pt, comsoc]{IEEEtran}
\IEEEoverridecommandlockouts
\usepackage{cite}
\usepackage{amsmath,amssymb,amsfonts}
\usepackage{algorithmic}
\usepackage{graphicx}
\usepackage{svg}
\usepackage{textcomp}
\usepackage{xcolor}
\def\BibTeX{{\rm B\kern-.05em{\sc i\kern-.025em b}\kern-.08em
    T\kern-.1667em\lower.7ex\hbox{E}\kern-.125emX}}
\begin{document}

\title{End-to-End Edge AI Service Provisioning Framework in 6G ORAN \\

\thanks{The authors are with the School of Aerospace, Transport and Manufacturing (SATM), Cranfield University, United Kingdom. The work is supported by EPSRC CHEDDAR: Communications Hub for Empowering Distributed clouD computing Applications and Research (EP/X040518/1) (EP/Y037421/1).}
}

\author{
\IEEEauthorblockN{Yun Tang}
\IEEEauthorblockA{yun.tang@cranfield.ac.uk}
\\
\IEEEauthorblockN{Obumneme Umealor}
\IEEEauthorblockA{obumneme.umealor@cranfield.ac.uk}
\and
\IEEEauthorblockN{Udhaya Chandhar Srinivasan}
\IEEEauthorblockA{u.srinivasan@cranfield.ac.uk}
\\
\IEEEauthorblockN{Dennis Kevogo}
\IEEEauthorblockA{Dennis.Kevogo@cranfield.ac.uk}
\and
\IEEEauthorblockN{Benjamin James Scott}
\IEEEauthorblockA{benjamin.scott@cranfield.ac.uk}
\\
\IEEEauthorblockN{Weisi Guo}
\IEEEauthorblockA{weisi.guo@cranfield.ac.uk}
}

\maketitle

\begin{abstract}
With the advent of 6G, Open Radio Access Network (O-RAN) architectures are evolving to support intelligent, adaptive, and automated network orchestration. This paper proposes a novel Edge AI and Network Service Orchestration framework that leverages Large Language Model (LLM) agents deployed as O-RAN rApps. The proposed LLM-agent-powered system enables interactive and intuitive orchestration by translating the user's use case description into deployable AI services and corresponding network configurations. The LLM agent automates multiple tasks, including AI model selection from repositories (e.g., Hugging Face), service deployment, network adaptation, and real-time monitoring via xApps. We implement a prototype using open-source O-RAN projects (OpenAirInterface and FlexRIC) to demonstrate the feasibility and functionality of our framework. Our demonstration showcases the end-to-end flow of AI service orchestration, from user interaction to network adaptation, ensuring Quality of Service (QoS) compliance. This work highlights the potential of integrating LLM-driven automation into 6G O-RAN ecosystems, paving the way for more accessible and efficient edge AI ecosystems.
\end{abstract}

\begin{IEEEkeywords}
Edge AI as a Service 6G, 6G, ORAN, LLM Agent
\end{IEEEkeywords}

\section{Introduction}

\begin{figure}[t]
    \centering
    \includegraphics[width=\columnwidth]{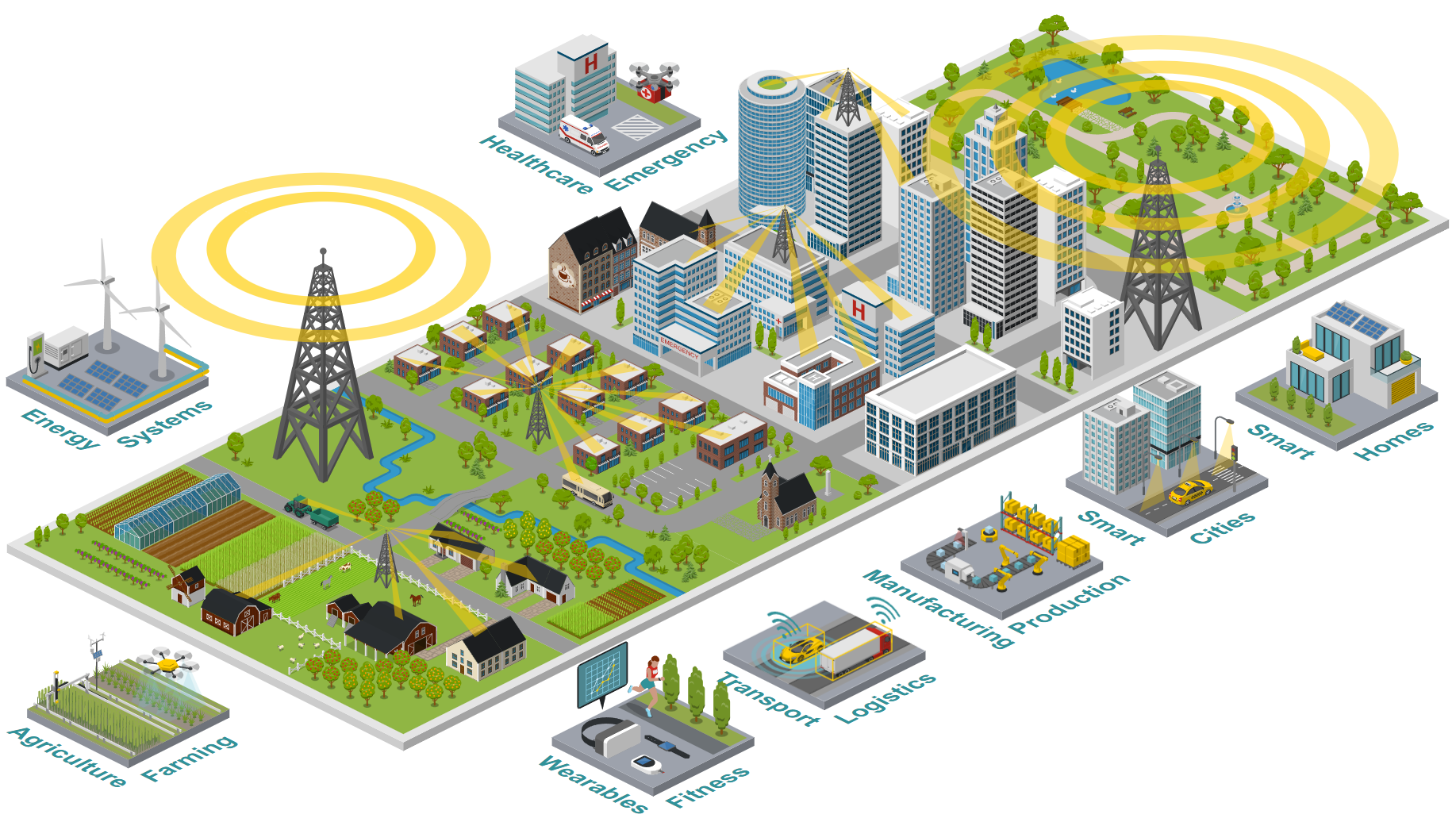}
    \caption{Edge AI use cases in 6G Network-powered modern society.}
    \label{fig:edge_ai_and_telecom}
\end{figure}

% Motivation for Edge AI services in 6G.
Edge intelligence is becoming increasingly important as next-generation use cases demand low-latency AI services close to end users (as shown in Fig.~\ref{fig:edge_ai_and_telecom}). However, deploying AI models at the network edge and configuring the underlying infrastructure remains complex and time-consuming. Without an automated solution, use case developers must manually integrate AI applications with distributed edge resources and confer with network service providers to tailor network settings for each use case, distracting from their primary goal of building innovative applications. Thus, there is a growing need for an end-to-end orchestration framework that abstracts these low-level tasks, allowing use case innovators to focus on application logic rather than the intricacies of AI service deployment and network configuration. 

In parallel, the evolution toward 6G networks is expected to embrace AI deeply into the network, for both network operations and the connected use cases \cite{letaief2019roadmap}. Research visions for 6G emphasize AI-native network management, where AI functions are integrated across cloud, core, and radio domains. For instance, the O-RAN architecture \cite{o-ran_alliance} already introduces RAN Intelligent Controllers (RICs) to enable data-driven control in the radio access network. Yet, current frameworks lack an end-to-end mechanism for orchestrating use case-facing AI services across domains in a holistic manner. 

Recent advances in large language models (LLMs) \cite{zhou2024large, long20246g} open up new possibilities: LLMs can interpret high-level intents, reason about complex tasks, and even generate configuration or code, which can be leveraged to automate network and service management.

Motivated by these trends, in this paper, we propose an LLM-powered end-to-end framework to streamline edge AI service deployment in 6G O-RAN. By using an LLM-based agent to handle user interaction, AI model selection, placement, network tuning and QoS monitoring, the framework enables developers to request and deploy edge AI services in an intent-based manner. The following sections detail the design and implementation of this framework and demonstrate how it can significantly simplify the deployment of edge AI applications while optimizing network behaviour end-to-end.

\section{Related Works}
\subsection{Edge AI Orchestration Approaches}
Prior works have explored platforms to orchestrate compute and AI workloads at the network edge. For example, Oakestra \cite{bartolomeo2023oakestra} is a lightweight hierarchical orchestrator for edge computing that tackles challenges like unreliable links and diverse hardware by federating resources and delegating tasks efficiently. Existing edge orchestration solutions primarily focus on container management, workload placement, and latency optimization. However, they often do not specifically address AI model management or integrated network control. Our work differs by concentrating on AI service orchestration in synergy with the networking service control, using an LLM-based agent to automate decisions typically made by network service operators.

% Brief introduction to O-RAN and its role in 6G networks.
\subsection{O-RAN and 6G}
The O-RAN initiative \cite{o-ran_alliance} specifies a disaggregated RAN architecture with open interfaces and intelligent controllers to enable innovation in network management. In O-RAN, a near-real-time RAN Intelligent Controller (near-RT RIC) hosts xApps that perform closed-loop control of the RAN (e.g., scheduling, slicing) on the order of 10 ms to 1 s, while a non-real-time RIC (within the Service Management and Orchestration framework) handles policy and analytics on $>1$s timescales. The RIC concept is a cornerstone that brings AI and automation into RAN operations. Early O-RAN deployments and research (often on 5G testbeds) have shown the feasibility of dynamic RAN optimization via xApps \cite{limani2024optimizing}. Looking ahead, AI in 6G will not only optimize the RAN but also coordinate end-to-end network behaviour across domains (i.e., RAN, edge, core, and cloud) \cite{li2024architecture}. This work builds on this background by using the O-RAN platform (RIC applications in RAN and service-based interfaces (SBI) in core) as the vehicle for network adaptation, under the guidance of a higher-level LLM-based orchestration agent.

% Existing research on Large Language Models (LLMs) in 6G automation.
\subsection{LLMs in 6G Networks}
The emergence of large language models has prompted exploration into their role within future network management and orchestration. For example, Maestro  \cite{Chatzistefanidis2024Maestro} proposes a framework where multiple LLM-based agents (representing multiple stakeholders such as network service providers) negotiate shared network resources. More generally, it is highlighted that LLMs can grasp user intent, reason about tasks, and execute commands, potentially redefining how we interact with and control network services \cite{abel2024large}. Our proposed framework can be seen as part of this emerging paradigm: it uses an LLM-based agent as an intelligent orchestrator that bridges the gap between a use case’s high-level intent and low-level network management and control actions.

\section{The Framework}

\begin{figure*}[t]
    \centering
    \includegraphics[width=\textwidth]{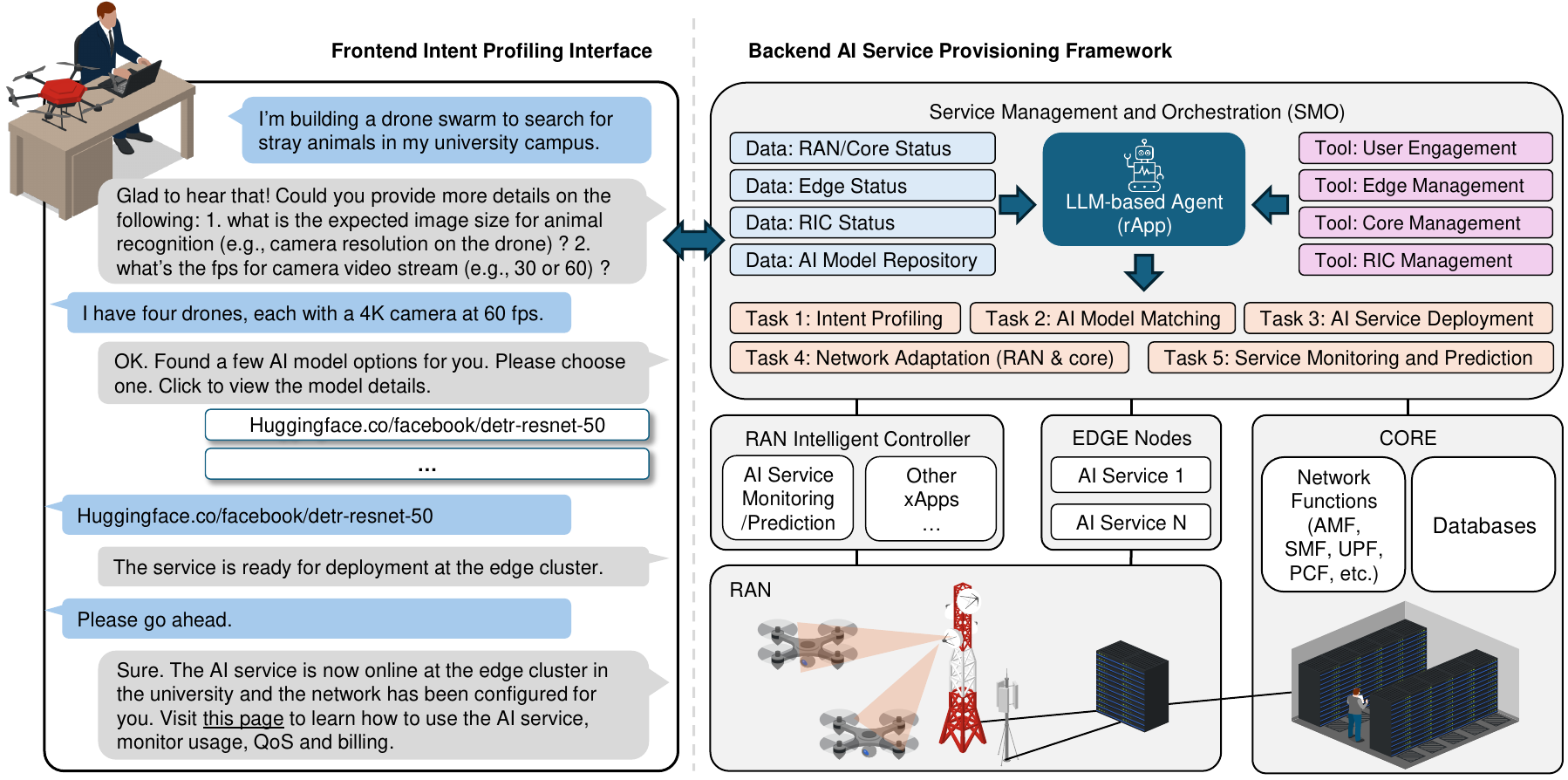}
    \caption{End-to-End Edge AI Service Provisioning Framework Overview.}
    \label{fig:framework-overview}
\end{figure*}

Fig~\ref{fig:framework-overview} presents an overview of the proposed framework. At the core of the framework is an LLM-based orchestrator agent that serves as the “brain” of the system. When a developer describes a use case (for example, ``I'm building a drone swarm to search for stray animals in ...''), the request is forwarded to the LLM agent. The agent is equipped with the necessary knowledge (from RAN, RIC, edge, core and AI model repository) and tools (user engagement and network management) to execute a predefined list of tasks including: selecting an appropriate AI model, determining a deployment strategy (which edge node to deploy on, what resources to allocate), and deciding on any network policy or RAN parameter adjustments needed. The LLM’s ability to perform reasoning over these decisions is what differentiates this framework from static rule-based orchestrators.

\subsection{AI Model Discovery and Selection}
One of the first tasks is discovering a suitable AI model that fulfils the use case's needs. The framework includes an AI model repository or registry which catalogues available pre-trained and edge-deployable models (and possibly training pipelines) for various AI tasks. Upon analysing the use case scenario, the LLM agent queries (via API or semantic matching) this repository to find models that match the described application. For instance, if the request is for an object detection service, the repository might have models like a trained ResNet or YOLO detector. The agent can consider factors such as the model’s accuracy (if the model is benchmarked against the developer's scenario), input requirements (camera feed, sensor data, etc.), and resource footprint when filtering the models. The output of this step is a list of chosen models and any metadata needed for deployment (model files, runtime requirements) and the developer is then promoted to make a selection. This task aims to abstract the complexity of AI model selection: instead of the developer having to search, choose and configure a model, the orchestrator’s intelligence handles it. In a more advanced implementation, this step could also involve model optimization – for example, choosing a quantized model for faster inference if accuracy is not the developer's primary concern.

\subsection{AI Service Deployment}
Once a model is selected, the framework proceeds to deploy the AI service on an appropriate edge computing node. The LLM agent is responsible for instantiating the AI service (e.g., as a containerized REST API or gRPC server) on a chosen site based on where the service can best run (e.g., at a radio unit site for ultra-low latency, or a regional edge data centre for heavier computing). 
% For our prototype, we assume a virtualized RAN and edge environment where components can be flexibly placed. 
The agent interacts (via the edge management tool) with an infrastructure manager (e.g., a Kubernetes cluster at the edge, an ETSI MEC platform, or a virtualization layer) to launch the AI service. It allocates necessary resources (vCPUs, memory, and GPU) to the AI container. Networking for the service is set up (e.g., via container networking or SDN controllers), ensuring the service is reachable from the RAN (to ingest data or serve results). At the end of this step, the requested AI functionality is running at the edge, but optimal QoS may not yet be achieved without adapting the network to this new load.

\subsection{Network Adaptation Mechanism}

Deploying an edge AI service can impose new demands on the network (for instance, high upstream data from cameras, or low-latency requirements for control signals). The framework addresses this by performing network adaptation in tandem with AI service deployment. Using the network management tools (represented by plum-coloured boxes in Fig~\ref{fig:framework-overview}), the LLM agent is provided with a list of permissible network control actions. It then determines which action to execute to meet the AI application’s QoS requirements. For example, if the AI service needs guaranteed bandwidth, the LLM might update the UE's network slice policies in the core network (e.g., using the SBI of Policy Control Function) and adjust the scheduling policies in RAN (using the O-RAN E2 interface) for the traffic of the corresponding network slice. This mechanism ensures an end-to-end approach: not only is the AI application deployed, but the connectivity and radio environment are tuned to facilitate its performance.

\subsection{Service Monitoring and Prediction}

To ensure that the deployed AI service meets the expected performance and network requirements, our framework includes a Service Monitoring and Prediction mechanism where the agent generates and deploys xApps on the fly to monitor AI service traffic and predict potential QoS issues, presenting real-time and predictive insights to the user.

\textbf{QoS Monitoring} Depending on the available service models or API endpoints, the key QoS metrics related to the AI service traffic can include:

\begin{itemize}
    \item \textbf{Latency}: End-to-end delay from the AI service request to response.
    \item \textbf{Throughput}: The bandwidth allocated and utilized for AI service communication.
    \item \textbf{Packet Loss}: Identifying network congestion that could degrade AI service quality.
    \item \textbf{Jitter}: Variability in packet transmission times that might affect real-time AI applications.
    \item \textbf{Service Availability}: Detecting disruptions in AI model inference or edge server downtime.
\end{itemize}

\textbf{Predictive Analytics}
Beyond real-time monitoring, xApps also integrate predictive analytics to anticipate potential QoS degradations before they impact the AI service. The xApps can employ:

\begin{itemize}
    \item \textbf{ML-based Traffic Prediction}: Historical QoS data is used to forecast future traffic trends and detect anomalies.

    \item \textbf{Threshold-based Alerts}: If a QoS parameter (e.g., latency exceeding a predefined threshold) signals potential degradation, proactive notifications are triggered.

    \item \textbf{AI Model Inference Delays}: The xApps can track the performance of the AI service itself such as inference times and compute resource usages.

\end{itemize}

\textbf{Reporting and Recommendations}
The monitoring xApps generate reports that are fed back to the user interface, which can include:

\begin{itemize}
    \item \textbf{Real-time QoS Dashboards} Graphical reports summarizing AI service performance and network conditions.

    \item \textbf{Proactive Notifications} Alerts when predicted QoS degradation is detected, allowing users to take preventive action.

    \item \textbf{Suggested Optimizations} Based on trends, the LLM agent may recommend actions such as selecting a different edge deployment location,  adjusting network configurations, or even providing explanations to manage user's expectations or adjust use case requirements.
    
\end{itemize}

This closed-loop feedback mechanism keeps users informed about the status of their requested AI service, offering transparency into network conditions and fostering greater trust in the network.

\section{Demonstration}

\subsection{Prototype Implementation}
Considering the diverse choices of RAN, RIC, core and edge solutions in the market with different interface and functionality implementations, we developed a prototype using widely adopted open-source platforms and conducted an end-to-end demonstration to showcase the feasibility of the envisioned framework. Specifically, we adopted OpenAirInterface (OAI) projects \cite{openairinterface} to simulate core network functions (\textit{oai-amf}, \textit{oai-smf}, \textit{oai-pcf} and \textit{oai-upf}), RAN (\textit{oai-gnb}), UEs (\textit{oai-nr-ue}), RIC (\textit{oai-flexric}), edge server (\textit{trf-gen-cn5g}) and cloud server (\textit{trf-gen-cn5g}). The LLM agent in our prototype is implemented using LangGraph \cite{langgraph} and powered by Google Gemini LLM (through API) \cite{gemini}. The AI model repository is the HuggingFace \cite{huggingface_models}. All components are containerized and orchestrated on a single testbed to emulate the edge and RAN nodes.

% While the scale is limited, the implementation covers the full cycle from input intent to networked AI service.

\subsection{End-to-End AI Service Provisioning Workflow}
% \textcolor{red}{YUN: below is a dummy demo scenario. the entire section is to be updated later based on what we can demo in reality before the deadline...}

\begin{figure*}
    \centering
    \includegraphics[width=\textwidth]{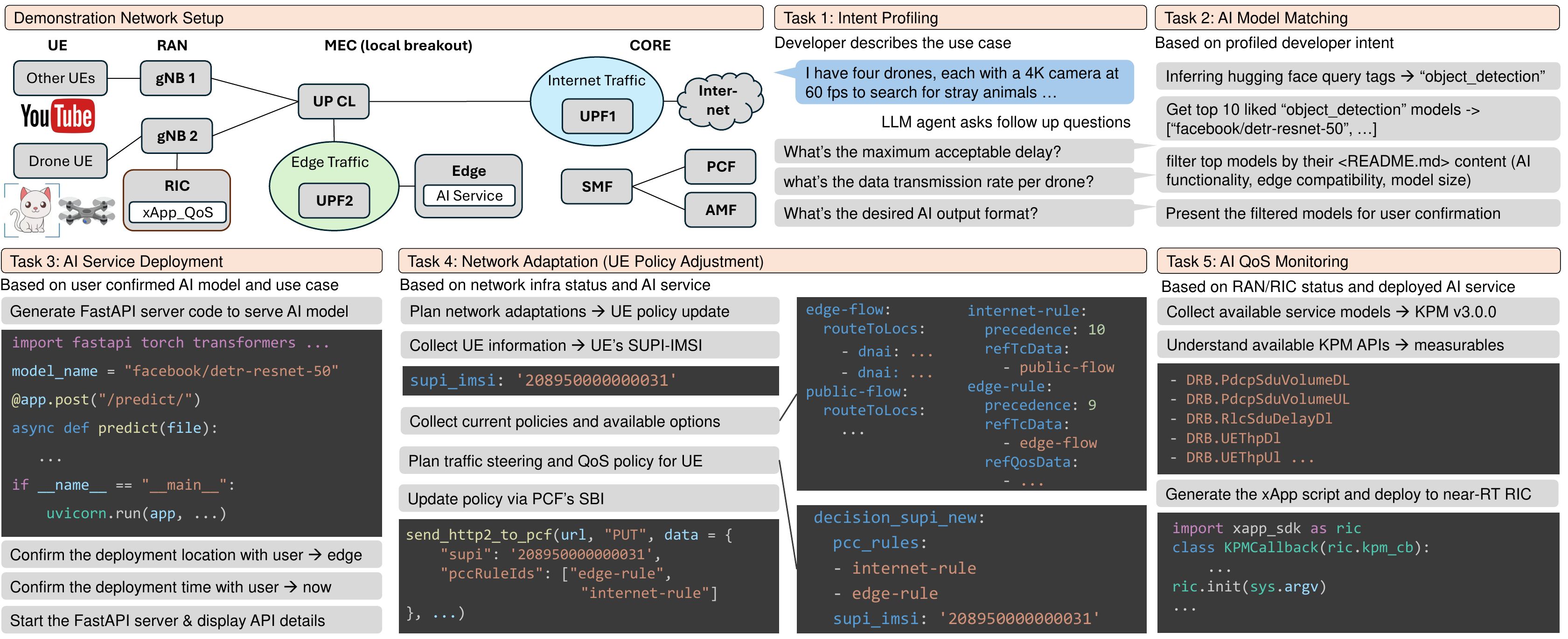}
    \caption{Agent tasks workflow for the demonstration use case.}
    \label{fig:demo-task-workflow}
\end{figure*}

As a proof-of-concept, we consider a use case where a developer wants to deploy a drone swarm to search for stray animals. The task results of the end-to-end workflow by our prototype setup are presented in Fig~\ref{fig:demo-task-workflow}:

\textbf{Intent Profiling} The LLM agent chats with the use case developer to collect as many requirements as possible.

\textbf{AI Model Matching} Given the use case of ``camera image-based stray animal searching,'' the agent initiates a search for the top 10 pre-trained AI models labelled as ``object\_detection'' on Hugging Face. For each retrieved model, the agent extracts and analyses the ``README.md'' file to assess its functionality and compatibility with edge deployment. For example, the agent evaluates whether the model can be seamlessly served through a REST API server. After filtering out incompatible models, the refined list is presented to the user for final selection.

\textbf{AI Service Deployment} After the user selects an AI model and confirms the suggested deployment parameters (target edge node and deployment time), the agent generates the necessary code to serve the AI model via a RESTful API or gRPC server. It then deploys and launches the server within the designated edge cluster. Upon successful deployment, the relevant API endpoint details are provided to the user for seamless access.

\textbf{Network Adaptation} Once the AI service is deployed and operational, the agent evaluates the available network adaptation options. Based on the current configuration of the RAN, MEC and CORE, the agent decides to update to traffic steering and QoS policies for the Drone UEs. To enforce these updates, the agent sends an API request to the Policy Control Function (PCF) via its Service-Based Interface (SBI). This ensures that when the drones connect to the network, their traffic is managed according to the newly established policies by the RAN and CORE components.

\textbf{AI QoS Monitoring} Finally, the LLM agent assesses the available RAN service models and the RIC SDK capabilities to implement the QoS monitoring of the newly deployed AI service. Specifically, the agent generates and deploys an xApp designed to continuously monitor and report QoS metrics from the RAN and, if applicable, the AI server in MEC.

In our lab test, a recorded video was used to simulate the camera input, and the AI service correctly identified objects of interest while maintaining seamless API responsiveness, which confirmed that the updated traffic policies ensured drone-edge communications. The entire workflow—from intent submission to AI service activation and network adaptation—was completed within minutes, demonstrating the efficiency of our LLM-powered framework. While this controlled test did not reveal any erroneous decisions, further validation is needed for real-world scenarios. Nonetheless, we believe the results validate the framework’s ability to automate and streamline edge AI service deployment and O-RAN network orchestration in an integrated and adaptive manner. More use case examples and source code are available at this link \footnote{https://github.com/orgs/Cranfield-GDP/repositories}.     

\section{Discussion}

While the proposed framework demonstrates the potential of LLM-driven orchestration in 6G O-RAN, several challenges remain. Below, we outline key limitations and potential future improvements:

\begin{itemize}
    \item \textbf{Lack of Real-time Validation} – The current framework executes LLM-generated actions (e.g., AI model selection, network adaptations, xApp deployment) without real-time verification, potentially leading to inefficiencies or suboptimal configurations.
    
    \textit{Future Work:} A validation loop using a network digital twin or simulation environment could pre-test decisions before deployment. Additionally, integrating telemetry feedback and iterative learning mechanisms could enable the LLM agent to refine its outputs dynamically.

    \item \textbf{Scalability and Request Prioritization} – When multiple users engage with the framework or when network resources reach full capacity, the lack of caching and prioritization can lead to service delays and inefficient resource allocation.
    
    \textit{Future Work:} Implementing a caching system for AI service requests and a priority-based scheduling mechanism could optimize resource utilization and improve response times, ensuring critical tasks receive preferential treatment.

    \item \textbf{Single-Agent Decision Bottleneck} – The framework currently relies on a single LLM agent for all orchestration tasks, which may limit scalability and introduce a single point of failure.
    
    \textit{Future Work:} A multi-agent approach, where specialized agents handle AI model selection, network optimization, and service monitoring separately, could improve reliability, parallel processing efficiency, and decision robustness.

\end{itemize}

By addressing these challenges, the framework can evolve into a more adaptive, scalable, and resilient solution for automated AI service deployment and network orchestration in 6G O-RAN.

\section{Conclusion}

This paper presented an LLM-powered orchestration framework to simplify edge AI service deployment in 6G O-RAN. The framework enables developers to describe their use case with high-level intents, with the LLM agent autonomously handling AI model selection, deployment, network adaptations and QoS monitoring. A prototype using OpenAirInterface and FlexRIC validated its feasibility, demonstrating efficient AI service orchestration with minimal manual intervention.

The results highlight the potential of integrating AI-driven reasoning with programmable networking interfaces to enable intelligent, adaptive networks. While promising, further research is needed to enhance real-time validation, request prioritization, and multi-agent decision-making. With these improvements, we believe the framework could play a key role in future AI-native 6G architectures, enabling scalable and autonomous edge intelligence.

\section*{Acknowledgment}
AI tools (ChatGPT, DeepSeek) have been used to revise (grammar and organization) author-written content.

\bibliographystyle{IEEEtran}
\bibliography{references}

\end{document}